\magnification \magstep 2
\vsize 240 true mm
\hsize 165 true mm

\centerline {\bf THE NONLINEAR-ELECTRODYNAMIC BENDING }
\centerline {\bf OF THE X-RAY AND GAMMA-RAY }
\centerline {\bf IN THE MAGNETIC FIELD OF  PULSARS AND MAGNETARS
\footnote{$^*$}{Published in Doklady Akademii Nauk, 2001, v. 380, p. 325,
in Russian. English version see in Doklady Physics}}

\centerline { Victor I. Denisov, Irene P. Denisova, Sergey I. Svertilov
\footnote{$^{**}$}{E - mail :\ \ Denisov@srd.sinp.msu.ru\ \
Sis@srdlan.npi.msu.su }}

\centerline {M.V. Lomonosov Moscow State University}
\centerline {Physics Department}

119899, Moscow, Vorob'evy Gory, MSU, Physics Department.

\centerline{ ABSTRACT}

It was shown that according to the non-linear
electrodynamics of vacuum electromagnetic rays should
bend in the field of magnetic dipole. The angles of ray
bending in the gravitational and magnetic fields of pulsars
and magnetars were obtained.
In the case of pulsars with $b\sim R\sim $ 100 km, $B_0\sim
10^{13}\ G$
the value of the angle of non-linear electrodynamic bending of a
ray in the Heisenberg-Euler theory will reach the value of
$\delta \psi_{NED}\sim 30'',$
and in the case of a magnetar with $B_0\sim 10^{15}\ G$
the angle $\delta \psi_{NED}$
will increase to $\delta \psi_{NED}\sim 1\ rad\sim 60^\circ .$
The angle of gravitational bending of a ray at neutron star
with $r_g$ = 3 km in the same conditions will be equal to
$\delta \psi_g\sim 0.06$ rad $\sim 4^\circ .$
 Observations can only be made in X-
rays and gamma-rays, for which the magnetosphere is quite
opaque.
 Because the distance from the Earth to the well-known
pulsars and magnetars is too large to observe the pure
effect of a ray bending.
The non-linear electrodynamic bending of a ray as well
as the gravitational bending will be revealed in the effect
of lensing.
\vfil
\eject

As it is well-known, the Maxwellian electrodynamics in the
absence of matter is a linear theory. Its predictions
concerning a very wide field of problems (except the
subatomic
level) are constantly confirmed with better and better
accuracy.
Quantum electrodynamics, which is based on Maxwell's
electrodynamics complemented by the renormalization
procedure,
also describes with good accuracy the various subatomic
processes, and, according to common opinion, is one of the
best
physical theories.

However, some fundamental physical reasons indicate that
electrodynamics is nonlinear not only in continuous medium,
but in vacuum also.

Several different models of non-linear electrodynamics of
vacuum are discussed now in the field theory. In the case of
a weak fields their Lagrangian can be written in the
parameterized form [1,2]:
$$ L={1\over 8\pi }\Big\{[{\bf  E}^2-{\bf  B}^2]+
\xi [\eta_1({\bf  E}^2-{\bf  B}^2)^2+
4\eta_2({\bf  B\   E})^2]\Big\}, \eqno(1)$$
where $\xi =1/B^2_q,$ $B_q=4.41\cdot 10^{13}$ $G,$
and the value of dimensionless post-Maxwellian parameters
$\eta_1$ ¨ $\eta_2$ depends on the choice of the model of non-
linear electrodynamics in vacuum.

In particular, the $\eta_1 $ and $\eta_2$
parameters in the Heisenberg-Euler non-linear
electrodynamics have quite concrete values
$\eta_1=\alpha /(45 \pi )=5.1\cdot 10^{-5}, \ \eta_2=7\alpha
/(180 \pi )=
9.0\cdot 10^{-5},$  while in the Born-Infeld theory they can
be expressed through the unknown constant
$a^2:$ $\eta_1=\eta_2=a^2B_q^2/4.$

The equations of electromagnetic field in the non-linear
electrodynamics are analogous to the equations of
electrodynamics of continuous medium:
$$curl\ {\bf  H} ={1 \over c} {\partial {\bf  D} \over
\partial t},\ \  \
div \ {\bf  D} =0, \ \ \
{\bf  D}=4\pi {\partial  L\over \partial {\bf
E}},$$
$$curl \ {\bf  E} =-{1 \over c} {\partial {\bf  B} \over
\partial t},\ \ \ div \ {\bf  B} =0,\ \ \
{\bf  H}=-4\pi {\partial  L\over \partial {\bf  B}}. $$
Using expression (1), it is not difficult to obtain the
expansion of vectors
${\bf  D}$ and ${\bf  H}$
in degrees $B/B_q$ and $E/B_q$ with the first order of
post-Mawellian accuracy:
$${\bf  D}={\bf  E}+2\xi \{\eta_1({\bf  E}^2-{\bf  B}^2){\bf
E}+2\eta_2
({\bf  B\  E}){\bf  B}\},$$
$${\bf  H}={\bf  B}+2\xi \{\eta_1({\bf  E}^2-{\bf  B}^2){\bf
B}-2\eta_2
({\bf  B\  E}){\bf  E}\}.$$
During a long time, the non-linear electrodynamics of vacuum
had no experimental confirmation and, hence, it seems
for many researchers as only abstract theoretical model. At
present its status has changed drastically. The experiments
on non-elastic scattering of laser photons on gamma-
quanta [3] confirm the non-linear character of
electrodynamics in vacuum. Thus, its different predictions
[4 -10], which can be tested experimentally, are of great
importance. However, non-linear corrections to the Maxwell
equations for the magnetic fields $B,\ E\sim 10^6$ G, which
can only be obtained in ground laboratories, are too
small to observe effects caused by them. On the other hand, some
objects are existing in nature, in which magnetic field induction
is comparable with the typical value of
magnetic field induction required for essential
manifestation of electrodynamics non-linearity in vacuum
$B\sim 4.41\cdot 10^{13}$ G, which was predicted by quantum
electrodynamics.

These objects are rotation-powered pulsars and magnetars.
Rotation-powered
pulsars are rapidly-rotating neutron stars with a strong
dipole magnetic field
$${\bf B}_0={3({\bf m \cdot  r}){\bf r}-r^2{\bf m}\over r^5},$$
where $\bf m$ is the magnetic dipole momentum.

Most of
them are the so-called "radio" pulsars, i.e. isolated neutron
stars, the emission of which is powered by the loss of
rotational energy [11]. The high sensitivity of
modern X-ray observatories permits to detect X-rays from
about several dozens of radio pulsars [12], but
only ten (or possibly six) rotation-powered pulsars were
observed at gamma-ray energies, i.e. so-called "gamma"-
pulsars [13]. Most of them have a rather
strong surface magnetic field ($max\ B_0 \geq 10^{12}$ G),
for
at least five gamma-pulsars the magnitude of $B_0$ exceeds
the
characteristic value of $ 4.41\cdot 10^{13}$ G [13].
Although, it is not yet clear how and where in the
pulsar magnetosphere the non-thermal high-energy emission
originates, several advanced models assume that particles
are accelerated above the neutron star surface and that
gamma-rays result from  curvature radiation or inverse
Compton induced pair cascade in a strong magnetic field near
the neutron star, i.e. the so-called "polar cap" models
[14-18].
These models give the opportunity to observe nonlinear
electrodynamics effects in gamma-pulsar hard emission.

As for magnetars, or strongly magnetized neutron stars
[19-22],
they are the most appropriate sources for manifestation of
nonlinear electrodynamics effects. It is supposed now that
magnetars can be revealed as soft gamma-ray repeaters (SGR)
mainly observed in hard X-rays and soft gamma rays [23]
and as so-called anomalous X-ray pulsars (AXP), i.e.
slow (6-12 s period) rotators mainly observed in the "classical"
X-ray range (2-20 keV) [24]. SGRs are transient
events characterized by brief ($\leq 1$ s) and relatively
soft
(peak photon energy $\sim 10-30$ keV) bursts of super-
Eddington
luminosity. There are four (or five) SGRs
known now. A possible connection between SGR and AXP was
suggested after the discovery of periodicities as well as
spin-down effect in some SGRs [25,26] with the values of
period and period derivative very similar to those of AXP.
Another similarity between SGR and AXP is based on the fact
that all of them appear to be associated with supernova
remnants (SNRs). The magnetic field $B$ magnitude in such
objects can be estimated from the period and period
derivative values, which lead to $max\ B_0\geq 10^{14}-
10^{15}$ G.
In most of the magnetar models the magnetic field is the
main energy source, powering both the persistent X-ray
emission and the soft gamma-ray burst activity.

Thus, non-linear electrodynamic effects will be much more
pronounced in the dipole magnetic fields of gamma-ray
pulsars and magnetars than in any ground laboratory
experiments.

The non-linear electrodynamic bending of electromagnetic
wave rays in the magnetic fields of pulsars and magnetars is
one of such effects. As it can be seen from simple
calculations [27], the bending angle $\delta \psi $ of a ray in
the gravitational and magnetic field of a neutron star depends on the
mutual orientation of a ray and the star magnetic dipole momentum
$\bf m$ as well as on the electromagnetic wave polarization.

In particular, for an electromagnetic wave, propagating in the
neutron star magnetic equator plane and polarized in the
same plane, the bending angle of a ray will be:
$$\delta \psi = {2r_g\over b}+{15\pi \eta_1\xi B_0^2R^6\over
4b^6},$$
where $r_g$ is the Schwartzhield radii of a star, $R$ is the
star geometrical radii,
$B_0$ is the induction of the magnetic field on the star magnetic
equator.

For an electromagnetic wave polarized normally to the magnetic
equator, one can obtain:
$$\delta \psi = {2r_g\over b}+{15\pi \eta_2\xi B_0^2R^6\over
4b^6}.$$
Thus, the angle of non-linear electromagnetic bending of a
ray
$\delta \psi_{NED} = 15\pi \eta_{1,2}\xi B_0^2R^6/(4 b^6)$
depends on the impact distance  $b$
differently than the angle of gravitational bending of a ray
$\delta \psi_g = 2r_g/ b.$
Besides, at $\eta_1\neq \eta_2$
the angle $\delta \psi_{NED}$ depends on the electromagnetic
wave polarization,
and the angle $\delta \psi_g$ does not. Thus, these
two parts of the total angle of ray bending can be
revealed by the processing of sufficiently complete
observational data, obtained for the different values of
impact distance $b$ and different polarizations of the
electromagnetic wave.

In the case of a pulsar with $b\sim R\sim $ 100 km, $B_0\sim
10^{13}\ G$
value of the angle of non-linear electrodynamic bending of a
ray in the Heisenberg-Euler theory will reach the value
$\delta \psi_{NED}\sim 30'',$
and in the case of a magnetar with $B_0\sim 10^{15}\ G$
the angle $\delta \psi_{NED}$
increases to $\delta \psi_{NED}\sim 1\ rad\sim 60^\circ .$

For comparison, we can indicate that the angle of
gravitational bending of a ray a neutron star with $r_g$ = 3
km under the same conditions will be equal
$\delta \psi_g\sim 0.06$ rad $\sim 4^\circ .$

However, it is rather difficult to observe this effect.
Firstly, both pulsars and magnetars have a sufficiently dense
magnetosphere. Thus, observations can be made only in the X-
rays and gamma-rays, for which the magnetosphere is quite
opaque.

Secondly, the distance from the Earth to the well-known
pulsars and magnetars is too large to observe the pure
effect of a ray bending

As our analysis shows [27], the non-linear electrodynamic
bending of a ray as well as the gravitational bending will
be revealed in the effect of lensing.
As a result, the external manifestations of effects of non-linear
electrodynamic and gravitational bending of a ray will
depend on the ratio of distances between the gamma-ray source,
pulsar or magnetar and the Earth. If the gamma-ray source is
extra-Galactic, then the scattering of emission flux by the
pulsar or magnetar gravitational and magnetic fields will be
large. Thus, the emission intensity, which underwent
significant bending of a ray, will be very small in the
vicinity of the Earth. In this case the effect of ray
bending will be revealed only as the harsh falling of detected
emission intensity even by the disappearingly small value of
the ray bending angles.

If gamma-rays were emitted near a magnetic neutron star
(for example, if it is part of a binary system, or the
gamma-ray sources are the regions, jointed to its surface),
then scattering of this emission flux by the gravitaional
and magnetic fields will not be so pronounced. In this case, although
the emission intensity detected near the Earth will decrease with
growing of the bending angle, it will not be so harsh as in the case
of the extra-Galactic source. Thus, for astrophysical objects
containing a neutron star with magnetic field at the level of $B\sim
10^{13}-10^{15}$ G, effects of non-linear and gravitational bending
of a ray will be assessible by observation even at the modern
accuracy level using the extra-Terrestrial astronomy technique.

\vskip 0.5 true cm

{\bf References:}

[1]. Denisov V.I., $\&$   Denisova I. P.,
    Optics and Spectroscopy,  2001, v. 90, p. 282.

[2]. Denisov V.I. $\&$  Denisova I. P.,
     Doklady Physics,  2001, v. 46, p. 377.

[3]. Burke  D.L.,  Field R.C., Horton-Smith G., Spencer J.E.,
        Walz D., Berridge S.C., Bugg W.M., Shmakov K., Weidemann A.W., Bula C.,
        McDolald K.T., Prebys E.J., Bamber C., Boege S.J., Koffas T.,
        Kotseroglou T.,
        Melissinos A.C., Meyerhofer D.D., Reis D.A., Ragg W. //
        Rhys. Rev. Lett. 1997. V.  79. P. 1626.

[4].  Alexandrov E.B., Anselm A.A. $\&$ Moskalev A.N.,
      JETP, 1985, v. 62, p. 680.

[5]. Rosanov N.N., JETP, 1993, v. 76, p. 991.

[6]. Bakalov D. et al  Quantum Semiclass. Opt., 1998, V. 10,
     {\rm \char 242} 1, p. 239-250.

[7]. Denisov  V.I. // Phys. Rev. 2000. V. D61  {\rm \char 242}  3. P. 036004.

[8]. Denisov V.I., Journal of Optics A: Pure and Applied Optics,
2000, V. 2, N 5, p. 372.

[9]. Rikken G.L.J.A. $\&$  Rizzo C., Phys. Rev. A,  2000,
     v. 63, p. 012107.

[10]. Denisov V.I. $\&$  Denisova I. P.,
      Optics and Spectroscopy, 2001, v. 90, p. 928.

[11]. Becker, W. 2000, Adv. Space Res. 25, 647.

[12]. Mereghetti, S. 2000, astro-ph$/$0102017.

[13]. Thomson, D.J. 2000, Adv. Space Res. 25, 659.

[14]. Sturrock, P.A. 1971, ApJ, 164, 529.

[15]. Ruderman, M.A., $\&$ Sutherland, P.G. 1975, ApJ, 196, 51.

[16]. Daugherty, J.K., $\&$ Harding, A.K. 1982, ApJ, 252, 337.

[17]. Daugherty, J.K.,  $\&$  Harding, A.K. 1996, ApJ, 458, 278.

[18].  Usov,  V.V., $\&$ Melrose, D.B. 1995, Aust.  J.  Phys.
       48, 571.

[19]. Zhang, B., $\&$ Harding, A.K. 2000, ApJ, 535, L51.

[20]. Duncan, R.C.,  $\&$  Thomson, C., 1992 ApJ 392, L9.

[21]. Thomson, C, $\&$ Duncan, R.C. 1995, MNRAS, 275, 255.

[22]. Thomson, C, $\&$ Duncan, R.C. 1996, ApJ, 473, 322.

[23]. Hurley, K. 2000, in Proc. 5th Huntsville GRB Symposium
      526 (New York: AIP), 763.

[24]. Gotthelf, E.V.,  $\&$  Vasisht, G. 1997, ApJ, 486, L133.

[25]. Hurley, K., et al. 1999, ApJ, 510, L111.

[26]. Kouveliotou, C., et al. 1998, Nature, 393, 235.

[27]. Denisov, V.I., Denisova, I. P. $\&$  Svertilov, S.I.,
      Doklady Akademii Nauk, 2001, v. 380, p. 325.

\bye